\documentclass[letterpaper, 10 pt, conference]{ieeeconf}  

\IEEEoverridecommandlockouts                              
\let\labelindent\relax
\usepackage{graphicx}
\usepackage[utf8]{inputenc}
\usepackage[caption=false]{subfig} 
\usepackage{multirow}
\usepackage{hyperref} 
\usepackage{amsmath}
\usepackage{enumitem}
\usepackage{cite}

\usepackage[ruled]{algorithm2e}
\newtheorem{definition}{Definition}

\title{\LARGE \bf
A Hybrid Approach To Hierarchical Density-based Cluster Selection
}

\author{Claudia Malzer$^{1}$ and Marcus Baum$^{2}$
\thanks{$^{1}$Claudia Malzer is with HAWK Hochschule f{\"u}r angewandte Wissenschaft und Kunst; Data Fusion Group at University of G{\"o}ttingen; and Max Planck Institute for Dynamics and Self-Organization in G{\"o}ttingen, Germany
        {\tt\small claudia.malzer@hawk.de}}%
\thanks{$^{2}$Marcus Baum is with Data Fusion Group, University of G{\"o}ttingen {\tt\small marcus.baum@cs.uni-goettingen.de}}%
}

\begin{document}

\maketitle
\thispagestyle{empty}
\pagestyle{empty}

\begin{abstract}

	HDBSCAN is a density-based clustering algorithm that constructs a cluster hierarchy tree and then uses a specific stability measure to extract flat clusters from the tree. We show how the application of an additional threshold value can result in a combination of DBSCAN* and HDBSCAN clusters, and demonstrate potential benefits of this hybrid approach when clustering data of variable densities. In particular, our approach is useful in scenarios where we require a low minimum cluster size but want to avoid an abundance of micro-clusters in high-density regions. The method can directly be applied to HDBSCAN's tree of cluster candidates and does not require any modifications to the hierarchy itself. It can  easily be integrated as an addition to existing HDBSCAN implementations.

\end{abstract}

\section{INTRODUCTION}

Clustering algorithms are used by researchers of various domains to explore and analyze patterns of similarity in their data. In \textit{density-based} clustering, clusters are regarded as data partitions that have a higher density than their surroundings, i.e. dense concentrations of objects are separated by areas of low density. Objects that do not meet a given density criterion are discarded as noise. Those kind of algorithms can be useful in many research fields, and are particularly well suited for spatial data mining \cite{review}, such as GPS coordinates or even radar reflections in traffic scenes.

In this paper, we discuss a problem that can occur in data sets with highly variable densities, especially when choosing a low minimum cluster size. In such case, we either completely miss some potentially relevant clusters, or receive a large number of small clusters in high-density regions that we would have intuitively regarded as only one or few clusters. We show how the application of an additional threshold value to the cluster hierarchy created by the algorithm HDBSCAN \cite{hdbscan} can provide a solution to this problem, and call this approach HDBSCAN($\hat{\epsilon}$). It can be viewed as a hybrid between DBSCAN* (see Section \ref{hdbscan_review}) and HDBSCAN in the sense that we select DBSCAN* clusters for a fixed threshold value, and HDBSCAN clusters from all data partitions not affected by the threshold.

Section \ref{related} provides a short overview of existing density-based clustering algorithms with focus on hierarchical solutions. Section \ref{hdbscan_review} gives a more detailed insight into HDBSCAN, since our approach is intended as an alternative -- or additional -- cluster selection method for the HDBSCAN hierarchy. In Section \ref{mymethod}, we first illustrate the mentioned problem on a real-life example and then formalize the application of a threshold to HDBSCAN. Results of our experiments with HDBSCAN($\hat{\epsilon}$) are presented and discussed in Section \ref{evaluation}. Section \ref{conclusions} concludes with a brief summary.

\section{Related Work}
\label{related}

The classic density-based algorithm DBSCAN \cite{dbscan} defines density as having a minimum number of objects (specified by an input parameter $minPts$) within the neighborhood of a certain radius. The size of the radius is specified by the distance threshold parameter $\epsilon$ (\textit{epsilon}). Objects that fulfill this density criterion are called \textit{core points}. In some cases, objects are no core points themselves but lie within the epsilon neighborhood of a core point. These are called \textit{border points} and are also assigned to clusters. All remaining points are classified as noise. 

DBSCAN's major weakness is that its epsilon parameter serves as a \textit{global} density threshold and it is therefore not possible to discover clusters of variable densities beyond epsilon. Many DBSCAN alternatives have been proposed with the aim of overcoming this problem. To name only two recent approaches, density peaks clustering by Rodriguez and Laio \cite{Rodriguez} and density-ratio based clustering by Zhu et al. \cite{densityratio} are algorithms designed to handle variable densities. For example, DBSCAN can be turned into a density-ratio based clustering approach by adding a third parameter. In density peaks clustering, a representation called \textit{decision graph} is used for manual selection of cluster centers.

However, the focus of our work lies on \textit{hierarchical} approaches with automatic cluster selection. The first hierarchical DBSCAN extension to become widely used was OPTICS by Ankerst et al. \cite{optics}. In contrast to DBSCAN, it constructs an ordered representation of the data set that allows to explore \textit{all} possible density levels. A value for the minimum cluster size is the only required input parameter. The authors also provided an automatic cluster selection method based on a parameter $\xi$. An alternative to this method was proposed by Sander et al. \cite{sander}. 

Dockhorn et al. \cite{alternateHdbscan} introduced $\epsilon$-HDBSCAN, where $minPts$ is gradually decreased for a DBSCAN hierarchy with fixed epsilon. They also proposed the \textit{edge quantile} method, which views the complete hierarchy as a cluster from which subclusters are cut off wherever the connecting edge length exceeds the 0.95 quantile of edge lengths \cite{VariableDBSCAN}. 

AUTO-HDS \cite{autohds} is another hierarchical method and quite similar to the newer but already widely used algorithm HDBSCAN by Campello et al. \cite{hdbscan}. However, it has been shown that HDBSCAN can outperform both AUTO-HDS and the combination of OPTICS with Sander et al.'s cluster extraction method \cite{hdbscan} \cite{hdbscan_acc}. HDBSCAN was proposed as an improved extension of DBSCAN and OPTICS for data exploration in diverse research fields. It has an efficient Python implementation \cite{hdbscan_python} that conforms to the \textit{scikit-learn} \cite{scikit-learn} library and supports a variety of metrics. The algorithm requires only a minimum cluster size as user input and then simplifies a complex single-linkage hierarchy to a smaller tree of candidate clusters. A flat solution is extracted automatically based on local cuts. In \cite{fosc}, Campello et al. introduce the ``Framework for Optimal Selection of Clusters'' (FOSC) that formalizes cluster selection through local cuts as an optimization problem. HDBSCAN's original selection approach is an example of a FOSC-compliant method.

The method proposed in this paper -- HDBSCAN($\hat{\epsilon}$) -- is intended as an alternative cluster selection approach for the HDBSCAN hierarchy. By implementing it in compliance to FOSC, it can easily be combined with other FOSC methods, and setting the threshold to 0 always results in the same clustering as the original HDBSCAN. We will give a brief overview of HDBSCAN and FOSC in the following section. 

	\section{The HDBSCAN Algorithm}
\label{hdbscan_review}

HDBSCAN is built on top of a slightly modified version of DBSCAN, called DBSCAN*, which declares border points as noise \cite{hdbscan}. Unlike DBSCAN(*), HDBSCAN does not select clusters based on a global epsilon threshold, but creates a hierarchy for all possible epsilon values with respect to $minPts$ as minimum cluster size.

\subsection{Mutual Reachability Distance}

In HDBSCAN, the \textit{core distance} $d\textsubscript{\textit{core}}$ is defined as the distance of an object to its \textit{minPts}-nearest neighbor. The constructed hierarchy is based on the \textit{mutual reachability distance}, which for two objects $x_{p}, x_{q}$ is $$max\{d_{core}(x_{p}), d_{core}(x_{q}), d(x_{p}, x_{q}) \}$$ where $d(x_{p}, x_{q})$ refers to the ``normal" distance according to the chosen metric, e.g. Euclidean distance. This approach separates sparse points from others by at least their core distance and makes the clustering more robust to noise.

The data set can then be represented as a graph with data objects as vertices, connected by weighted edges with the mutual reachability distances as weights. Using this graph to construct a minimum spanning tree and sorting its edges by mutual reachability distance results in a hierarchical tree structure (\textit{dendrogram}). By choosing an epsilon as a global horizontal cut value and selecting all clusters with at least $minPts$ points at this density level, we could retrieve the DBSCAN* clusters for this epsilon from the hierarchy.

\subsection{Condensed Cluster Hierarchy}

Since HDBSCAN aims at discovering clusters of variable densities, it instead proceeds to building a simplified version of the complex hierarchy tree, the \textit{condensed cluster tree}. This approach follows the \textit{tree pruning} concept of which many variants exist in literature, such as \textit{runt pruning} by Stuetzle \cite{stuetzle2003} or pruning a tree created by a robust single linkage algorithm by Chaudhuri et al. \cite{chaudhuri}. A detailed explanation of the relationships between those approaches and HDBSCAN is provided by McInnes and Healy \cite{hdbscan_acc}.

Starting from the root, HDBSCAN regards each cluster split as a \textit{true} split only if both child clusters contain at least $minPts$ objects. If they both contain less than $minPts$ objects, the cluster is considered as having disappeared at this density level. If only one of the children has less than $minPts$ objects, the interpretation is that the parent cluster has simply lost points but still exists. ``Lost'' points are regarded as noise. This simplification process results in a hierarchy of candidate clusters at different density levels.

\subsection{Stability-based Cluster Selection}
\label{hdbscan_selection}
Given the condensed cluster tree, one possibility is to simply select all leaf nodes. These are the clusters with lowest epsilon values in the hierarchy, and represent clusters that cannot be split up any further with respect to $minPts$. This selection method is one of two provided options in HDBSCAN's Python implementation -- from now on called HDBSCAN(leaf) -- and results in very fine-grained clusters. 

The other option is \textit{eom}, short for \textit{excess of mass}. This method is recommended by Campello et al. \cite{hdbscan} as the optimal global solution to the problem of finding clusters with the highest \textit{stability}, which they define as 
\begin{equation}
\label{eq:1}
\begin{aligned}
S(C_{i}) &= \sum_{x_{j} \in C_{i}} ( \lambda_{max}(x_{j}, C_{i}) - \lambda_{min}(C_{i}))\\  
&= \sum_{x_{j} \in C_{i}} ( \frac{1}{\epsilon_{min}(x_{j}, C_{i})} - \frac{1}{\epsilon_{max}(C_{i})})
\end{aligned}
\end{equation}

where the density value $\lambda$ is simply set to $\frac{1}{\epsilon}$. This means that $\lambda$ values become larger from root towards leaves, whereas the corresponding $\epsilon$ distance values become smaller. Subtracting $\lambda_{min}(C_{i})$, which corresponds to the density level at which cluster $C_{i}$ first appears, from the value beyond which object $x_{j} \in C_{i}$ no longer belongs to $C_{i}$, results in a measure of \textit{lifetime} for $x_{j}$. The sum of all object lifetimes within $C_{i}$ leads to the overall cluster lifetime $S(C_{i})$, which is called \textit{stability} because the clusters with longest lifetimes are considered to be the most stable ones. 

The authors then formalize an optimization problem for maximizing the sum of these cluster stabilities as  

\begin{equation}
\label{eq:2}
\begin{aligned}
&	\max_{\delta_{2}, ..., \delta_{k}}  J = \sum_{i=2}^{k} \delta_{i}S(\textbf{C}_{i}) 
\\
	&\text{subject to }
	\begin{cases}
	
	\delta_{i} \in \{0, 1\}, i = 2, ..., k   
	\\
	\sum_{j \in \textbf{I}_{h}} \delta_{j} = 1, \forall h \in \textbf{L}
	\end{cases}
\end{aligned}
\end{equation}
with \textbf{L} = $\{ h | \textbf{C}_{h} \text{is leaf cluster}\}$ as leaves, $\textbf{I}_{h}$ as the set of clusters on the paths from leaves to the excluded root, and $\delta_{i}$ as boolean indicator whether the respective cluster is selected or not. The definition ensures that no more than one cluster can be selected on any branch from leaf towards root.

To solve this problem, HDBSCAN's selection algorithm traverses the cluster tree bottom-up. The stability value of each node is compared to the sum of stability values of its nested subclusters. In this way, stabilities are propagated and updated when going up the tree until the cluster with highest stability is found and selected on each branch.

\subsection{Framework for Optimal Selection of Clusters}

The excess of mass (eom) method for cluster extraction as explained above conforms to the generic ``Framework for Optimal Selection of Clusters'' (FOSC) introduced by Campello et al. \cite{fosc}. FOSC requires two essential properties: First, the chosen measure for cluster selection -- in this case, the stability criterion -- must be \textit{local}, i.e. it can be computed for each cluster independently of other selected clusters. Second, it must be \textit{additive}, i.e. it must be meaningful to add up the computed cluster values (in this case, cluster stabilities) so that an optimization problem as shown above can be formulated as maximizing the sum of values. Moreover, it must be ensured that exactly one cluster is selected on each branch. This formal problem can then be solved by traversing the hierarchy tree bottom-up starting from the leaves, deciding for each candidate cluster whether it should be part of the final solution. FOSC thus provides an efficient way of finding the globally optimal solution for the extraction of clusters according to the chosen measure.

\section{HDBSCAN($\hat{\epsilon}$): A Threshold for Cluster Splits}
\label{mymethod}
In this section, we introduce an approach for selecting clusters from the HDBSCAN hierarchy based on a distance threshold $\hat{\epsilon}$. Our motivation for this approach is given below, followed by a formal definition.

\subsection{Motivation}
\label{motivation}

HDBSCAN is a powerful clustering algorithm for unsupervised data exploration. However, for some applications, the single input parameter $minPts$ might not be sufficient to discover the clusters that best represent the underlying data structure. In particular, let us consider a large data set distributed such that there is a high number of very dense objects in some areas, and only few objects in other areas. If we were only interested in the highly populated areas, HDBSCAN would give us good results for a $minPts$ value large enough to declare sparse regions as noise and dense regions as clusters. In some cases, however, we do not want all observations in sparse environments to be marked as noise: these areas might naturally contain fewer objects, but small yet dense groups of objects that do exist might be just as relevant as the ones in regions with lots of data.

\begin{figure}
	
		\centering
	\subfloat[The sample data set]{%
		\fbox{\includegraphics[width=0.23\textwidth]{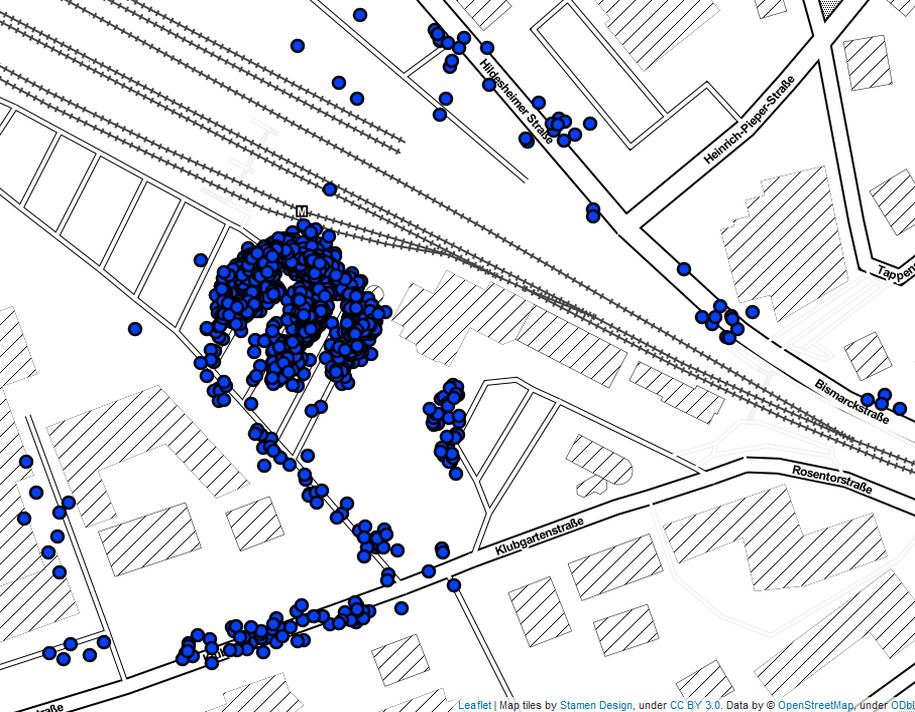}}
		\label{cluster_data}
		}  
		\subfloat[HDBSCAN with $minPts$ = 4]{%
		\fbox{\includegraphics[width=0.23\textwidth]{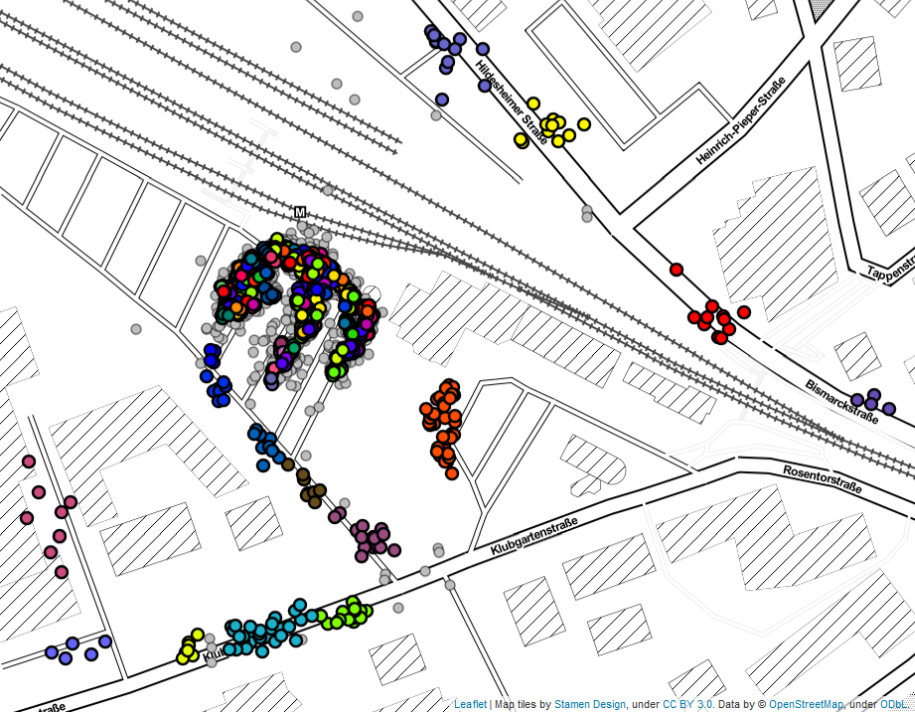}}
		\label{hdbscan_demo1}
		}

		
		\centering
		\captionsetup{justification=centering}
		\subfloat[OPTICS with $minPts$ = 4,\\ $\xi$ = 0.05]{%
		\fbox{	
		\includegraphics[width=0.22\textwidth]{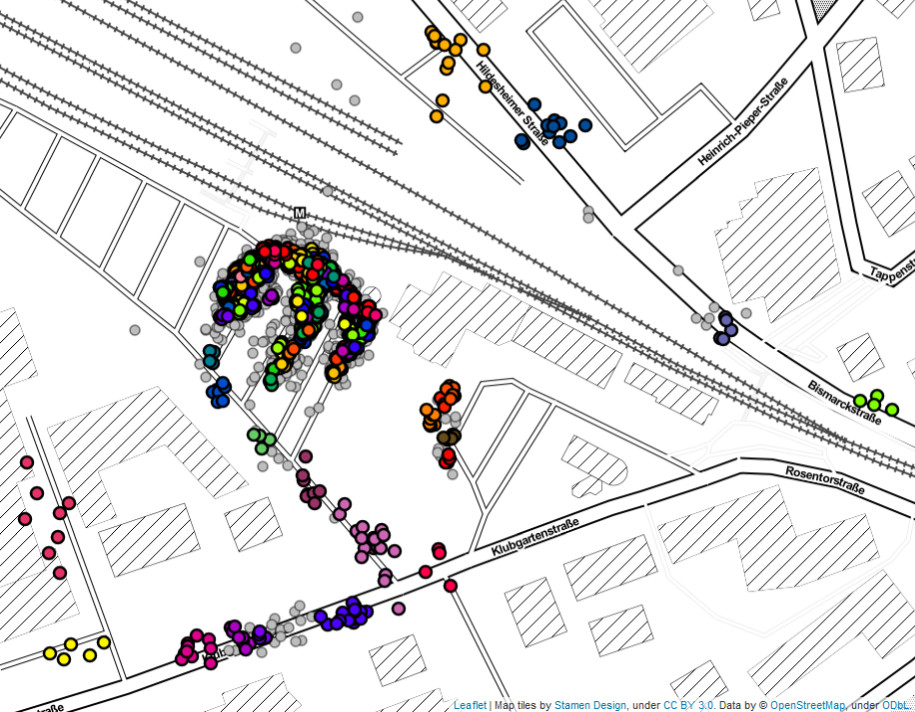}
		}
		\label{optics_demo1}
	}
	\subfloat[DBSCAN* with $minPts$ = 4, \\$\epsilon$ = 5 meters]{%
		\fbox{\includegraphics[width=0.22\textwidth]{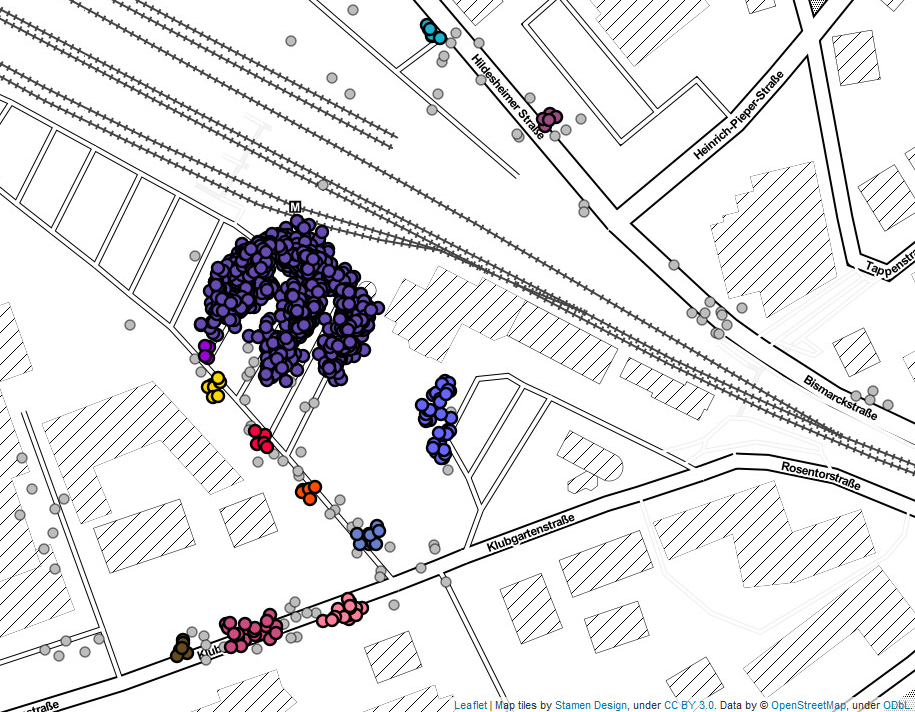}}
		\label{dbscan_demo1}
	}
	\caption{Clustering results for a sample data set of GPS locations. 
	}
\label{datasets}
\end{figure}

Figure \ref{cluster_data} demonstrates such a scenario. It shows around 2800 GPS data points on a map extract, representing recorded pick-up and drop-off locations from a door-to-door demand-responsive ride pooling system \cite{ecobus}. Our aim was to assign addresses requested by customers to the closest areas where ride pooling vehicles were actually able to stop in the past, i.e. in compliance with traffic rules and available space. The largest (visual) data cluster can be found around the train station. Smaller clusters are placed along the streets, depending on the requested location. Since we are considering a door-to-door system where customers are not bound to collective pick-up or drop-off locations, even small groups of 4 or 5 points are of interest to us.  

\setlength{\textfloatsep}{1\baselineskip plus 0.2\baselineskip minus 0.5\baselineskip}

Figure \ref{hdbscan_demo1} presents the clustering result with HDBSCAN's default selection method \textit{eom}, from now on referred to as HDBSCAN(eom). Using $minPts$ = 4, the algorithm successfully discovers all the small clusters while declaring obvious outliers or groups with less than 4 points as noise (gray points). However, in the high-density area around the train station, it generates a very large number of micro-clusters. In our case, this is not what we want: we would prefer one or only few clusters representing the location. This would be possible by increasing $minPts$, but at the cost of losing small clusters in less dense areas or merging them into other clusters separated by a relatively large distance.

We clustered the same data set with OPTICS from \textit{scikit-learn} and DBSCAN* from HDBSCAN's Python implementation. The $minPts$ parameter was set to 4 in both cases. For \mbox{OPTICS}, we tried $\xi$ values between 0.03 to 0.05, and for DBSCAN*, we tried $\epsilon$ values between 3 and 10 meters (haversine distance). In each case, we chose a value that intuitively produces the best results. Figure \ref{optics_demo1} shows OPTICS clusters for $\xi$ = 0.05, which are similar to HDBSCAN. Figure \ref{dbscan_demo1} depicts DBSCAN* results for $\epsilon$ = 5 meters, which seems clearly better suitable for our application. However, DBSCAN* neglects potentially important clusters with densities beyond the chosen threshold. In particular, this applies to the two groups of objects on the bottom-left and another two on the far-right. A larger $\epsilon$ would cover these objects, but at the same time merge other clusters. Note that this demonstration is based on only a small sample. Applying DBSCAN(*) to a larger data set increases its tendency to single-linkage effects when increasing $\epsilon$, such as extending the largest cluster down the streets.

For a scenario like this, a \textit{combination} between HDBSCAN and DBSCAN* would be useful. The essential idea is that instead of performing a cut through the entire hierarchy, we just prevent clusters below a given threshold from splitting up any further. We could then still select regular HDBSCAN clusters from data partitions not affected by the threshold. All others would be DBSCAN* clusters.

\subsection{Formal Definition}
\label{formaldefinition}
We define the application of a threshold to the HDBSCAN hierarchy as a selection method in accordance with the framework FOSC \cite{fosc} as explained earlier. To formalize our cluster selection criterion, we introduce the notions of \textit{epsilon stable} and \textit{epsilon stability}. 

\begin{definition}[Epsilon stable]
	\label{definition1}A cluster $C_{i}$ with $i = \{2, ..., k\}$ is called \textit{epsilon stable} if  $\epsilon_{max}(C_{i}) > \hat{\epsilon}$ for a given $\hat{\epsilon} > 0$.
\end{definition}

As explained earlier, $\lambda_{min}(C_{i}) = \frac{1}{\epsilon_{max}(C_{i})}$ is the density level at which cluster $C_{i}$ appears. Note that this is equal to the level at which it split off its parent cluster. Hence we call a cluster epsilon stable if the split from its parents occurred at a distance above our threshold $\hat{\epsilon}$ (or below the level $\lambda_{min}(C_{i})$, respectively) and formally define epsilon stability as follows: 

\begin{definition}[Epsilon Stability]
	\label{definition2}
	\[		
	ES(C_{i}) =
	\begin{cases}
	
	\lambda_{min}(C_{i})  & \text{if $C_{i}$ is epsilon stable}
	\\
	0 & \text{otherwise}
	\end{cases}
	\]
	
	\label{epsilon_stability}
\end{definition}

If we select the cluster with highest epsilon stability on each path of the HDBSCAN condensed hierarchy tree, we end up with all the clusters that we do not want to split up any further w.r.t. $\hat{\epsilon}$ and $minPts$. Their parents split up at some distance $\epsilon_{min} > \hat{\epsilon}$, which is equal to the $\epsilon_{max}$ value on the level where their children appear. While those children are still valid clusters, they are not allowed to split up themselves because either they are leaf clusters or the level $\lambda_{max} = \frac{1}{\epsilon_{min}}$ at which the split would happen is above the threshold.

Given these definitions, we can formulate the optimization problem that maximizes the sum of epsilon stabilities in the same way as previously shown in Section \ref{hdbscan_selection} (Problem \ref{eq:2}), with the only difference that we replace $S({C}_{i})$ with $ES({C}_{i})$. The definition ensures that we extract only the maximum epsilon stable cluster on each path from leaf towards root.

\subsection{Selection Algorithm}

The original selection algorithm by Campello et al. \cite{hdbscan} takes into account that the total stability value needs to be updated at each step. In our case, this is not necessary due to the definition of epsilon stability. The pseudo code in Algorithm \ref{alg1} thus shows a simplified version of the original code that has been adapted to our problem. First, we mark all leaves of the \mbox{HDBSCAN} cluster hierarchy as selected. If a leaf has previously been marked as not being a cluster, or if it is already epsilon stable (i.e., $\lambda_{min} <= \frac{1}{\hat{\epsilon}}$ for input parameter $\hat{\epsilon}$), we continue with the next leaf. Otherwise, we traverse upwards until we find an ascendant that split off its parent at a density level $\lambda_{min} <= \frac{1}{\hat{\epsilon}}$. If we find one before reaching the root, we select it as a cluster and unselect all of its descendants.

\begin{algorithm}

	\caption{Solution to the Optimization Problem}\label{alg1}
	\vspace*{1.2mm}
	\begin{enumerate}[leftmargin=0.5cm,label*=\arabic*.]
		\item Initialize $\delta_{(.)} = 1$ for all leaves
		\item Do bottom-up from all leaves (excluding the root):
		\begin{enumerate}[leftmargin=*,label*=\arabic*.]
			\item If $ES(C_{i}) > 0$ or $\delta_{C_{i}} = 0$, continue to next leaf
			\item Else if  $ES(C_{i}) = 0$ and $ES(C_{PARENT(i)}) > 0$, set $\delta_{PARENT(i)} = 1$ and set $\delta_{(.)} = 0$ for all \newline nodes in $C_{PARENT(i)}$'s subtree
		\end{enumerate}
	\end{enumerate}
\end{algorithm}

\begin{figure}
\vspace*{-0.5cm}
	\centering
	\subfloat[Hierarchy with $\lambda_{min}$ \newline values per level]{%
		\includegraphics[width=0.19\textwidth]{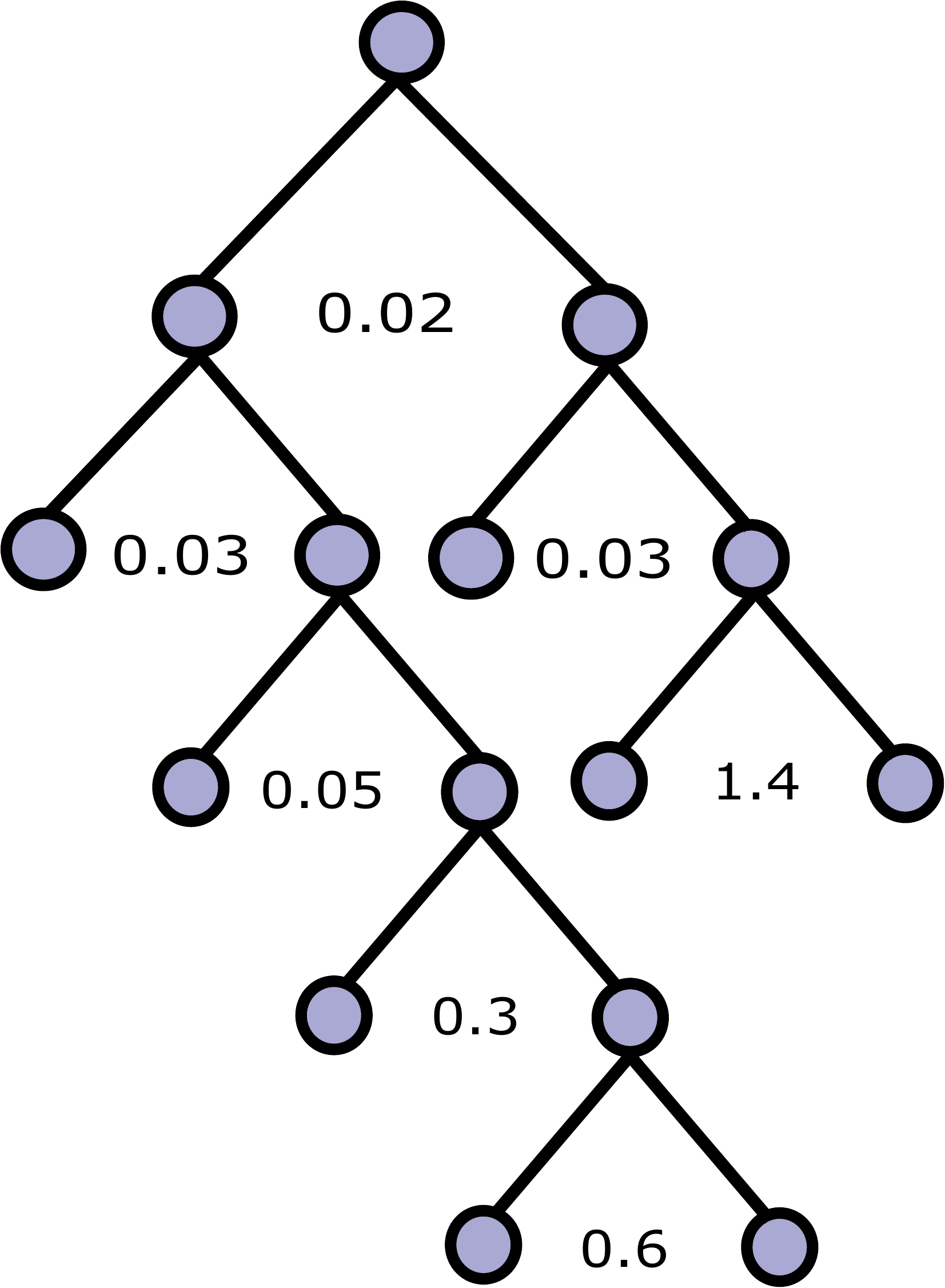}
		\label{fig:sub1}
	} \hspace{0.05\textwidth}
	\subfloat[Hierarchy with $ES$ values for $\lambda = 0.2$]{%
		\includegraphics[width=0.19\textwidth]{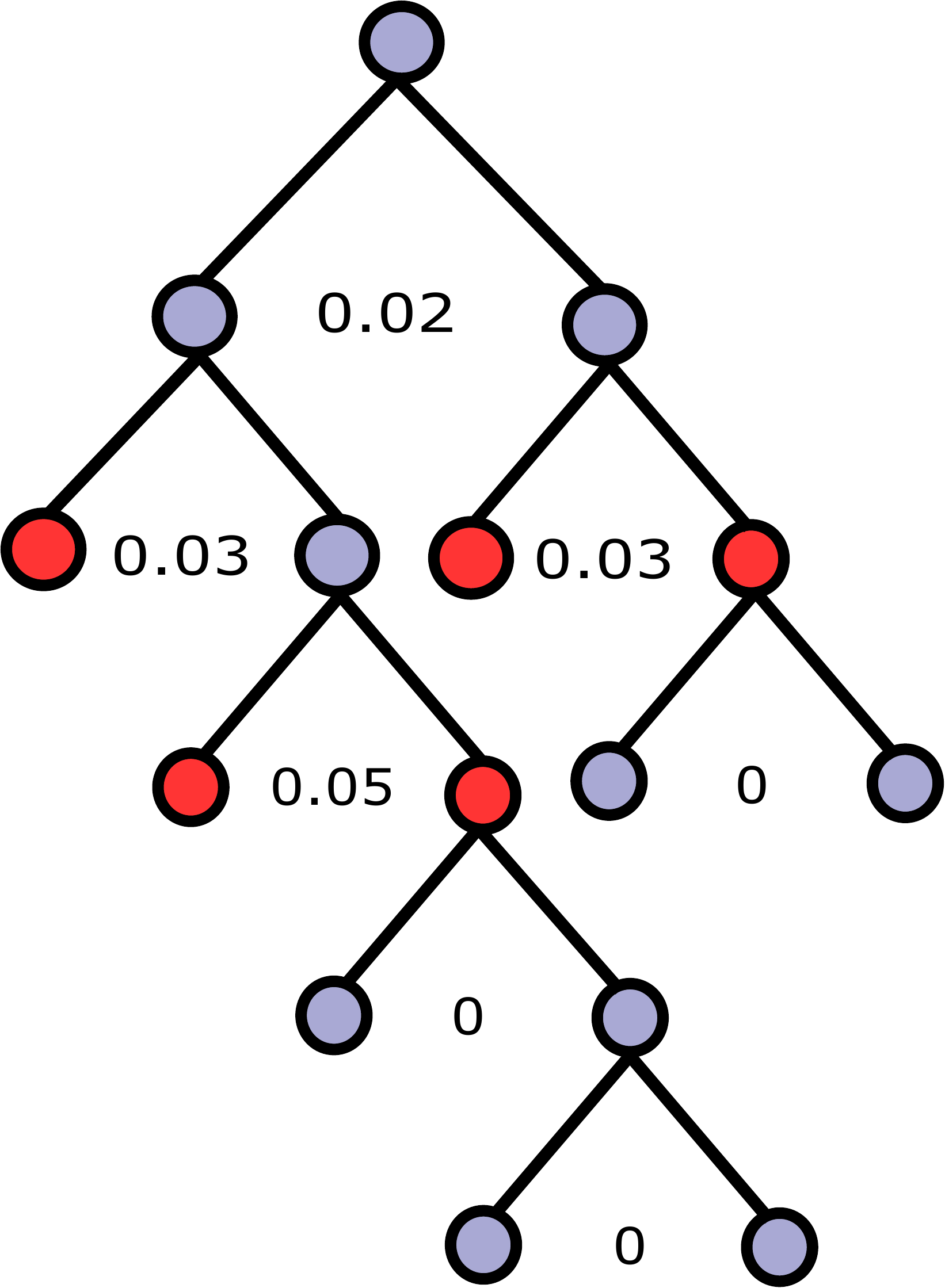} 
		\label{fig:sub2}
	} 

	\caption{The cluster tree in Figure \ref{fig:sub1} is annotated with the $\lambda_{min}$ value per hierarchy level. Figure \ref{fig:sub2} shows the same tree annotated with epsilon stability values for a given threshold $\hat{\epsilon} = 5$ (or $\lambda = 0.2$, respectively). On each path, the cluster with maximum epsilon stability is highlighted in red. }
	\label{clustertree}
\end{figure}
\setlength{\textfloatsep}{1\baselineskip plus 0.2\baselineskip minus 0.5\baselineskip}

Since $ES(C_{i})$ decreases as we traverse up the tree (unless for 0 values), this procedure ensures that we are selecting the maximum epsilon stable cluster on each path. This is equal to selecting the first cluster on the path that ``was born" at a distance greater than $\hat{\epsilon}$ and is thus not allowed to split up.  

The concept is illustrated in Figure \ref{clustertree}, which shows the cluster tree for a small sample data set. The tree on the left is annotated with $\lambda_{min}$ values. On the right, the same tree is annotated with corresponding epsilon stability values for $\hat{\epsilon} = 5$ meters, or $\lambda = \frac{1}{\hat{\epsilon}} = 0.2$. According to Definition \ref{epsilon_stability}, each cluster on a level that is not epsilon stable is set to 0. All others receive their $\lambda_{min}$ as epsilon stability value. 

It can be seen that the density levels with $\lambda_{min}$ values of 0.6, 0.3 and 1.4 exceed the threshold of 0.2. This indicates that the parents of clusters on these levels split up at a distance lower than 5 meters. By starting from the leaves and selecting the ascendant with maximum epsilon stability value on each path, we receive the final set of clusters (highlighted in red). Note that instead of starting from the leaves, we could optionally also start with the nodes selected by HDBSCAN(eom). This would combine both methods, i.e. after selecting the most stable clusters we could still decide to merge clusters up to the given threshold value.

HDBSCAN($\hat{\epsilon}$) can be viewed as FOSC-compliant: the ES measure is \textit{local} in the sense that for each cluster, we can decide whether to set the value to 0 or to $\lambda_{min}$ independently of the values of other clusters, and since we can add up the values for our optimization problem, it is also \textit{additive}.

\section{Experiments and Discussion}
\label{evaluation}

\begin{figure}

	\centering
	\subfloat[DS1]{%
		\includegraphics[width=0.24\textwidth]{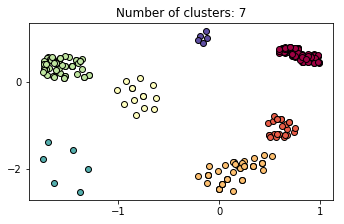}
		\label{fig:dataset_sub1}
	}
	\subfloat[DS2]{%
		\includegraphics[width=0.215\textwidth]{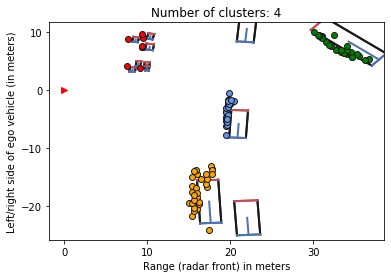}
		\label{fig:dataset_sub2}
	}
	\caption{A synthetic data set (left) and a real data set (right).}

	\label{datasets}
\end{figure}

Before applying HDBSCAN($\hat{\epsilon}$) to the GPS data set introduced in Section \ref{motivation}, we also demonstrate the algorithm on a synthetic data set and an alternative real data set as illustrated in Figure \ref{datasets}. Different (random) colors represent the true labels. Both data sets contain clusters of variable shapes and densities and serve as examples for cases where HDBSCAN(eom) alone leads to an abundance of micro-clusters. For comparison, each data set was also clustered with OPTICS and DBSCAN*. Since we focus on scenarios where we require a low minimum cluster size, $minPts$ was set to 4 for DS1 and 2 for DS2. Epsilon in DBSCAN* and HDBSCAN($\hat{\epsilon}$) was in each case set to a value that results in the most accurate clustering w.r.t. the ground truth data. OPTICS' $\xi$ value was set to 0.05 in both cases. Lower or higher values did not improve the result.

\begin{figure}
	
	\begin{minipage}[b]{.4\textwidth}
		\centering
		\subfloat[HDBSCAN(eom)]{%
			\includegraphics[width=0.55\textwidth]{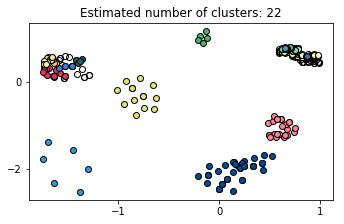}
			\label{fig:ds2_hdbscan}
		}
		\subfloat[OPTICS with $\xi = 0.05$]{%
			\includegraphics[width=0.55\textwidth]{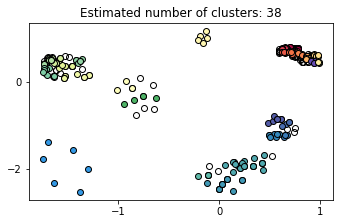}
			\label{fig:ds2_optics}
		}
	\end{minipage}
	
	\begin{minipage}[b]{.4\textwidth}
		\subfloat[DBSCAN* with $\epsilon = 0.34$]{%
			\includegraphics[width=0.55\textwidth]{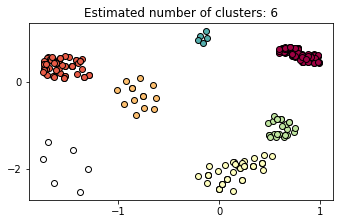}
			\label{fig:ds2_dbscan}
		}
		\subfloat[HDBSCAN($\hat{\epsilon}$) with $\hat{\epsilon} = 0.1$]{%
			\includegraphics[width=0.55\textwidth]{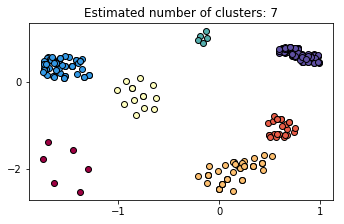}
			\label{fig:ds2_hdbscan_e}
		}
		
		\caption{Clustering results for DS1. Colorless points are noise.}
		\label{results_d1}
	\end{minipage}
\end{figure}

 Figure \ref{results_d1} shows that \mbox{HDBSCAN(eom)} discovers clusters of variable densities in DS1, but breaks up high-density regions into micro-clusters. OPTICS results are no improvement over HDBSCAN(eom). DBSCAN* almost achieves the desired output, but we could not find an epsilon value that recognizes the cluster on the bottom-left while keeping the remaining clusters separated. Figure \ref{fig:ds2_dbscan} shows the result for $\epsilon = 0.34$, which is just large enough not to merge clusters and at the same time not declaring any more points as noise. 

\begin{figure}
	
	\begin{minipage}[b]{.4\textwidth}
	\centering
	\subfloat[HDBSCAN(eom)]{%
		\includegraphics[width=0.55\textwidth]{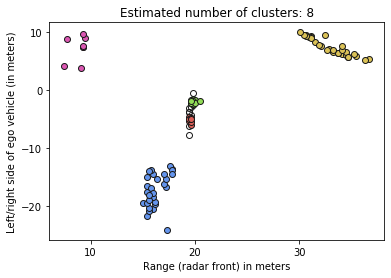}
		\label{fig:ds1_hdbscan}
	}
	\subfloat[OPTICS with $\xi = 0.05$]{%
		\includegraphics[width=0.55\textwidth]{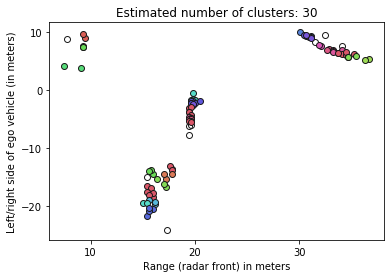}
		\label{fig:ds1_optics}
	}
		\end{minipage}
	
		\begin{minipage}[b]{.4\textwidth}
				\subfloat[DBSCAN* with $\epsilon = 0.13$]{%
				\includegraphics[width=0.55\textwidth]{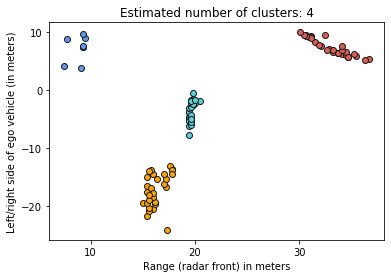}
				\label{fig:ds1_dbscan}
			}
			\subfloat[HDBSCAN($\hat{\epsilon}$) with $\hat{\epsilon} = 0.1$]{%
				\includegraphics[width=0.55\textwidth]{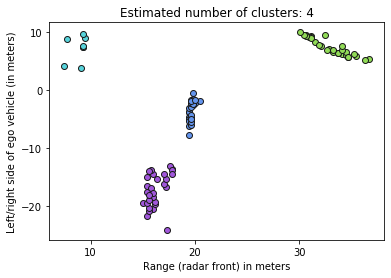}
				\label{fig:ds1_hdbscan_e}
			}
			\end{minipage}
		\caption{Clustering results for DS2. Colorless points are noise.}
		\label{results_d2}
		\end{figure}

DS2 is based on radar data from \textit{nuscenes} \cite{nuscenes} (scene-0553) with reflections from pedestrians and vehicles. In Figure \ref{fig:dataset_sub2}, the ego vehicle position is marked with a red arrow, and bounding boxes for moving objects are included. Note that the manually chosen ``true'' labels consider the group of pedestrians (red points) as one cluster, but results that detect subgroups would also be acceptable. Potential clutter and stationary objects below a threshold of 0.1 $\frac{m}{s}$ were removed. By clustering based on a 3D data set where Doppler velocity and Cartesian x/y coordinates were scaled to values between 0 and 1, we used a very simple radar clustering approach. However, the example again demonstrates that for a low $minPts$, OPTICS and HDBSCAN(eom) might create undesired micro-clusters (see Figure \ref{results_d2}). For DBSCAN*, an epsilon lower than 0.13 neglects relevant reflections from the pedestrian group. For larger values, a perfect result can be achieved. However, larger epsilons increase the risk of merging other clusters, e.g. vehicles driving side by side. In contrast, HDBSCAN($\hat{\epsilon}$) does not require an $\hat{\epsilon}$ larger than 0.1. Overall, the range of possible epsilon values for the best result are 0.34 - 0.38 (DS1) and 0.13 - 0.35 (DS2) for DBSCAN* and 0.1 - 0.38 (DS1) and 0.04 - 0.35 (DS2) for HDBSCAN($\hat{\epsilon}$). This indicates that HDBSCAN($\hat{\epsilon}$) is less sensitive to the choice of its epsilon parameter.

Results are represented by the Adjusted Rand Index (ARI) as seen in Table \ref{results}. Note that the ARI, a commonly used cluster validation measure, does not consider noise. Hence, DBSCAN* and HDBSCAN($\hat{\epsilon}$) both achieve a perfect ARI score of 1 for DS1, although DBSCAN* marks the bottom-left cluster as noise. For DS2, we recorded DBSCAN*'s perfect score for $\epsilon = 0.13$ instead of the result for $\epsilon = \hat{\epsilon}$.

\begin{table}
	\vspace{ 2mm}
	\centering
	\renewcommand{\arraystretch}{1.2}
	\caption{Clustering results for data set DS1 and DS2 in terms of Adjusted Rand Index (ARI) and percentage of data points not marked as noise (\%c). HDB. is short for HDBSCAN. }
	\begin{tabular}{p{0.08\columnwidth}|c|c|c|c|c|c|c|c}
		\hline
		\multirow{2}{0.08\columnwidth}{\centering\textbf{Data \\Set}} & 
		\multicolumn{2}{c|}{\textbf{HDB. (eom)}} &
		
		\multicolumn{2}{c|}{\textbf{OPTICS}} & 
		\multicolumn{2}{c|}{\textbf{DBSCAN*}} & 
		\multicolumn{2}{c}{\textbf{HDB. ($\hat{\epsilon}$)}} \\
		
		\cline{2-9}
		& \textbf{ARI} & \textbf{\%c} & \textbf{ARI} & \textbf{\%c} 
		& \textbf{ARI} & \textbf{\%c} & \textbf{ARI} & \textbf{\%c} 
		\\
		
		\hline
		\centering DS1 & 0.28 & 0.75 & 0.11 & 0.78 & 1 & 0.98 &  1 &  1 \\ \hline
		\centering DS2 & 0.80 & 0.92 & 0.09 & 0.90 & 1 & 1 &  1 & 1\\ \hline
	
	\end{tabular}
	
	\label{results}
\end{table} 

\begin{figure}
	\centering
	\setlength{\fboxsep}{0pt}%
	\setlength{\fboxrule}{1pt}%
	\fbox{
		\includegraphics[width=0.32\textwidth]{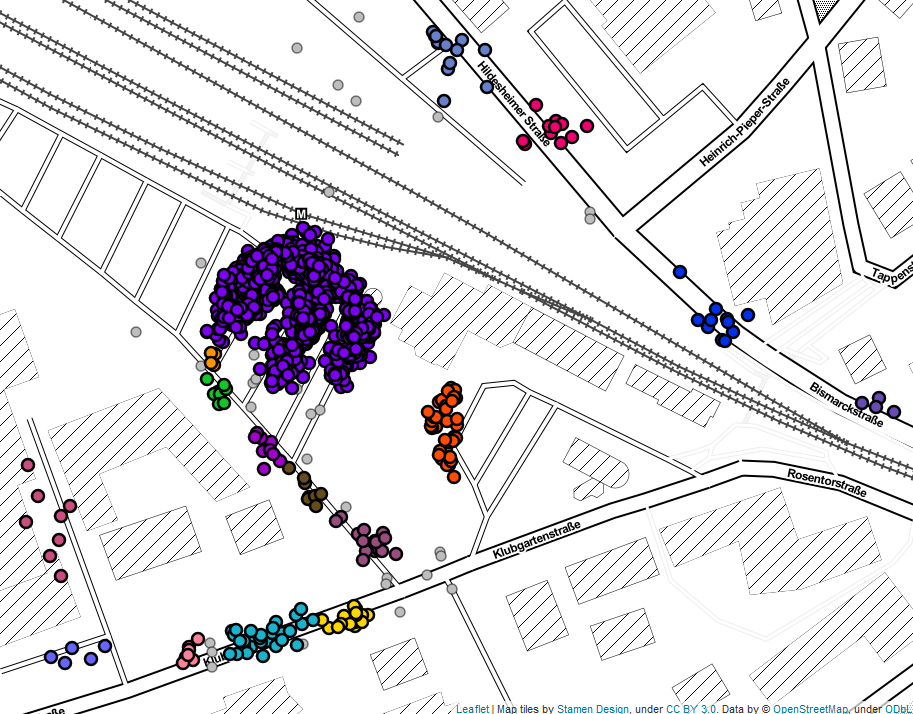} 
	}
	\caption{HDBSCAN($\hat{\epsilon}$) with $minPts = 4$, $\hat{\epsilon} = 5$ meters}
	\label{hdbscan_e_demo1}
\end{figure}
\setlength{\textfloatsep}{1\baselineskip plus 0.2\baselineskip minus 0.5\baselineskip}
Finally, we applied HDBSCAN($\hat{\epsilon}$) to the sample of GPS data introduced in Section \ref{motivation}. Figure \ref{hdbscan_e_demo1} shows the result for $\hat{\epsilon} = 5$ meters. Compared to DBSCAN* in Figure \ref{dbscan_demo1} with the same epsilon value, and \mbox{HDBSCAN(eom)} in Figure \ref{hdbscan_demo1}, we notice that we indeed receive a combination of both. We no longer lose clusters of variable densities beyond the given epsilon, but avoid the high number of micro-clusters in the original clustering, which was an undesired side-effect of having to choose a low $minPts$ value. Note that for the given parameter setting, running HDBSCAN($\hat{\epsilon}$) based on \textit{eom} or \textit{leaf} would not make any difference: $\hat{\epsilon}$ neutralizes the effect of HDBSCAN's stability calculation. For a lower threshold, e.g. $\hat{\epsilon}$ = 3, some minor differences can be noticed.

In general, the most suitable $\hat{\epsilon}$ value is certainly not always easy to choose. For GPS data, it is quite intuitive to decide on a distance threshold, but in higher-dimensional data, this becomes more difficult. Another limitation of our method is that cutting the hierarchy at a fixed threshold can neglect meaningful subclusters. Hence, it inherits DBSCAN(*)'s shortcomings wherever it uses a fixed value to select clusters. 
\addtolength{\textheight}{-12cm} 
It could also be argued that the threshold could already be applied when building the hierarchy and does not require the definition of an optimization problem or FOSC-compatibility. However, the application at selection stage has the major advantage that we do not need to modify HDBSCAN's original hierarchy and then re-build it every time we apply a different threshold. Instead, we can work with the original tree (or even a cache) and then efficiently explore our data set with different selection methods and thresholds. It has been shown that FOSC cluster extraction does not increase the overall complexity of HDBSCAN, which is $O(n^{2})$ \cite{fosc}.

\section{Summary and Conclusion}
\label{conclusions}

We introduced the cluster selection method HDBSCAN($\hat{\epsilon}$) that applies a distance threshold to HDBSCAN's hierarchy and therefore acts like a hybrid between DBSCAN* and HDBSCAN: for data partitions affected by the given threshold $\hat{\epsilon}$, we extract DBSCAN* results, for all others \mbox{HDBSCAN} clusters. The method is designed to be compatible to the framework FOSC and can be combined with other FOSC-compliant methods. It can easily be integrated into existing HDBSCAN implementations and is already available as part of the scikit-learn compatible Python module \cite{hdbscan_python}. 

We belief that this extension will prove to be valuable particularly in clustering spatial data, but it could also be applied to different kind of data. Future work might include combinations with further selection methods such as semi-supervised approaches. 



\bibliographystyle{IEEEtran}


\end{document}